%% file: main.tex
\documentclass{svproc}

\usepackage{array} 
\usepackage{booktabs} 
\usepackage{tabularx} 
\usepackage{multirow} 

	\usepackage[utf8]{inputenc} 
	\usepackage[T1]{fontenc} 
	\usepackage[english]{babel} 
	
	\usepackage{xcolor}
	
	\usepackage{amsmath,amssymb}
	\usepackage{xfrac} 
	
	\usepackage{enumitem}

	\usepackage{caption}
	\usepackage{subcaption}
	
	\usepackage{graphicx} 
	
	\usepackage{multicol} 
	\usepackage{footmisc} 
	
	\usepackage{siunitx}
	\usepackage[english=american]{csquotes}
	\usepackage[nonumberlist,acronym,nomain]{glossaries}
	
	\loadglsentries{acronyms.tex}
	
	\usepackage{url}
	
	\usepackage[hidelinks]{hyperref} 
	
	\usepackage[capitalise]{cleveref} 
	
	\newcommand{\eg}{e.\,g.\ } 
	\newcommand{\ie}{i.\,e.\ }
	\newcommand{\cf}{cf.\ }
	\newcolumntype{L}[1]{>{\raggedright\let\newline\\\arraybackslash\hspace{0pt}}p{#1}}
	\newcolumntype{C}[1]{>{\centering\let\newline\\\arraybackslash\hspace{0pt}}p{#1}}
	\newcolumntype{R}[1]{>{\raggedleft\let\newline\\\arraybackslash\hspace{0pt}}p{#1}}
	
	\usepackage[final,
	todonotes={obeyFinal,english,}]{changes}
	\definechangesauthor[name={Stephan}, color=orange]{sl}
	
\begin{document}
		\mainmatter              
		%
		\title{Reference Network and Localization Architecture for Smart Manufacturing based on 5G}
		%
		\titlerunning{Reference Network and Localization Architecture for Smart Manufacturing}

		\author{Stephan Ludwig\inst{1} \and Doris Aschenbrenner\inst{1}\inst{2} \and Marvin Schürle\inst{1}\inst{3}\and Henrik Klessig\inst{4} \and Michael Karrenbauer\inst{5} \and Huanzhuo Wu\inst{6} \and Maroua Taghouti\inst{7} \and Pedro Lozano\inst{8} \and Hans D.~Schotten\inst{5} \and Frank H.~P.~Fitzek\inst{6}}
		\authorrunning{Stephan Ludwig et al.} 
		%
		\tocauthor{Stephan Ludwig, Doris Aschenbrenner, Henrik Klessig, Michael Karrenbauer, Huanzhuo Wu, Maroua~Taghouti, Pedro Lozano, Hans D.~Schotten and Frank H.~P.~Fitzek}
		\institute{Aalen University, 73430 Aalen, Germany,
			\email{Stephan.Ludwig@hs-aalen.de}
			\and
			TU Delft, Faculty of Industrial Design Engineering, 2628CE Delft, The Netherlands
			\and
			Conclurer GmbH, 89522 Heidenheim an der Brenz, Germany
			\and
			Robert Bosch GmbH -- Corporate Research, 71272 Renningen, Germany
			\and
			Technische Universität Kaiserslautern, 67663 Kaiserslautern, Germany
			\and
			Technische Universität Dresden, 01062 Dresden, Germany
			\and
			Technische Universität Berlin, 10623 Berlin, Germany
			\and
			Ericsson GmbH, 52134 Herzogenrath, Germany
		}
		
		\maketitle              
		
		\begin{abstract}
			5G promises to shift Industry 4.0 to the next level by allowing flexible production.
			However, many communication standards are used throughout a production site, which will stay so in the foreseeable future.
			Furthermore, localization of assets will be equally valuable in order to get to a higher level of automation.
			This paper proposes a reference architecture for a convergent localization and communication network for smart manufacturing that combines 5G with other existing technologies and focuses on high-mix low-volume application, in particular at small and medium-sized enterprises. 
			The architecture is derived from a set of functional requirements, and we describe different views on this architecture to show how the requirements can be fulfilled. 
			It connects private and public mobile networks with local networking technologies to achieve a flexible setup addressing many industrial use cases.
			
			
			
	\end{abstract}
	
	\section{Introduction}\label{sec_intro}
	Industry~4.0 \cite{spath2013produktionsarbeit}, 
or the \gls{iiot}, will merge physical and real worlds as connected \glspl{cpps} that will enable an unprecedented production efficiency and degree of automation, as well as highly flexible and reconfigurable \enquote{smart} manufacturing---including quality management. 
Flexibility and changeability became the main enablers of staying competitive in global markets. 
Especially \glspl{sme} have played an essential role in supplying large enterprises with (customized) parts for products/intermediate goods, which are produced in small batch-sizes in a \gls{hmlv} production. 
In that, efficient retooling and the optimization of setting-up times is mandatory, 
as this currently takes around \SI{1.8}{\hour} of lost manufacturing time \cite{conrad2016single}. 
Additionally, precise real-time localization of goods is foreseen to significantly cost-optimize intra-logistics.
It is expected that the 5th generation of mobile communications (5G), together with its impact on the network infrastructure, will become a key factor therein. 
While an architecture for 5G 
has been developed by 3GPP, smart manufacturing use cases have been proposed in the 5G ACIA in \cite{5g_alliance_for_connected_industries_and_automation_5g_acia_5g_2018}, whose requirements led to 3GPP study items in preparation for 3GPP Rel-16 and subsequent.
Localization with precision below \SI{1}{\m} are planned for Rel-18 (finished by mid of 2024).

Despite these developments, including private networks using dedicated spectrum locally on their premises, 5G's introduction faces hurdles and challenges, which often relate to the problem of \textit{How to (securely) integrate a 5G network into a (existing) manufacturing environment with all its peculiarities?}
In order to address the challenges, 
reference architectures have been published in the literature: 
The \gls{iira} \cite{lin_IndustrialInternetThings_2019} vertically integrates production technologies in the \gls{iiot}. 
It contains different views with respect to different dimensions: 
Functional domains, system characteristics and crosscutting functions;  
and explicitly names connectivity as such a crosscutting function. 
Complementary to the \gls{iira}, the \gls{rami} \cite{adolphs_ReferenceArchitectureModel_2015} addresses three different dimensions related to higher-level layers, including communications, the life cycle value stream and hierarchy levels, the latter essentially resembling the automation pyramid. 
However, existing architectures define only few regarding the integration of operational \gls{ot} and \gls{ict} like 5G into an industrial network, for which interdisciplinary joint work from different fields like manufacturing, automation, manufacturing IT, IT security and communication technology is required.
Although there are already promising approaches to use 5G to enable more flexible (and more dynamic) manufacturing processes \cite{cheng2018industrial}, 
mainly large companies carry out these examples and target large production lines. 
Nevertheless, there are still huge differences between \glspl{sme} and large enterprises \cite{spena2016requirements,MASOOD2020103261}, 
specifically regarding new technological approaches. 

Consequently, we propose a 5G location-based context-aware production \enquote{small cell}.
Compared to existing applications of 5G in manufacturing, our approach aims at using flexible 5G small cells---without full area coverage, but only where needed---that federate among each other. 
In \cref{sec_requ}, we help create a mutual understanding by summarizing interdisciplinary requirements, especially from a plant's manager's perspective.
Based on \cite{karrenbauer2019future}, we propose a reference network and localization architecture in \cref{sec_overview} with 5G as its nucleus and specifically targeting \gls{hmlv} needs, which fulfills the requirements and integrates well into the \gls{iira} and \glsentryshort{rami} (cf. \cite{karrenbauer2019future}). 
The 5G small cell is developed into a heterogeneous industrial communication and localization network, 
expanding the components-based architecture in \cite{karrenbauer2019future,ludwig_5G_2018} 
by virtualization, data flow, security (leveraging strong 5G security features \cite{5gamericas_security_2018}) and management views.
This enables a dynamically configurable constellation on the shopfloor, representing the reality of \glspl{sme} with \gls{hmlv} manufacturing requirements.
In the last \cref{sec_summary}, we summarize this article and outline further necessary research steps.

\section{Summary of Functional System Requirements}\label{sec_requ}
When plant managers consider a new system, they weigh out the economic benefits with monetary effort (return on invest), while assessing how technical requirements are fulfilled. 
In most cases, there is no \emph{one and only} killer use case that pays off the invest. 
Furthermore, a smooth transition from legacy solutions to new technology is necessary, requiring interoperability across systems. 
We focus on related requirements in this section originating from in-field experience collected by the 5GANG project \cite{schildtknecht}.

\textit{Req-A) Interoperability and Network/Localization Convergence:} 
A new system shall support use cases with a common and interoperable infrastructure, ideally deployable worldwide---while it is understood that \gls{ue} implements only subsets of the standard and hence supports not all technical requirements of all use cases simultaneously. 
Especially \gls{sme} hesitate to spend the significant costs for a full coverage 5G network on their production premises.
Flexible small cells could be used just there, where connectivity is required, and multiple small cells shall act as a federated system with non-disruptive handover.
Since plant managers have a plethora of existing networking technologies already, another requirement relates to network convergence. 
The system shall transparently integrate new devices without adjustments for incremental retrofitting, which also applies for different existing wired or wireless access technologies like Industrial and classic Ethernet, field bus networks, WiFi, Bluetooth and \gls{lpwan} (LoRa, mioty etc.). 
Although 5G allows end devices to specialize into application specific need using \gls{embb}, \gls{urllc} and \gls{mmtc}, while a base station provides all three flavors, applications can have more strict requirements than 5G might fulfill, \eg in battery lifetime, where (ultra) \gls{lpwan} technologies like LoRa or mioty might outperform 5G \gls{mmtc}.
With this and gateway support, devices establish multiple connections to different base stations for load balancing, seamless handover, extended coverage, mesh networks, 5G device-to-device communication and increased reliability. 
Besides communication, localization support of assets enables an optimization of the intra-logistics.
Hence, future systems will have to support a heterogeneous set of localization technologies on the physical layer such that they converge with communications.
In addition, this 5G++ flexible small cell (\enquote{5G++ FlexiCell}) shall integrate into the production network and optimize the \gls{e2e} data transfer performance.

\textit{Req-B) Flexible Deployment and Operator Models:} 
Plant managers require flexible network deployment, ownership and operator models, which facilitate easy and fast deployment, operation and management. 
This needs to be ensured in \glspl{sme}, which do not afford owning network infrastructure, 
as well as in large-size companies, which might want to operate their own network. 
Independent of the operator model, sensitive production data (which is the key value and know-how of production) shall stay on-premise, reducing the risk of data breach. 
To ensure global applicability and flexibility, private networks shall be supported in licensed and unlicensed bands 
and in MNO-owned or dedicated, local spectrum. 
Finally, indoor and outdoor operation needs to be supported, \eg for inbound logistics use cases. 

\textit{Req-C) Manageability:} 
Infrastructure personnel might not always be available in \glspl{sme}, or may be centralized in large companies. 
Therefore, automatic device integration and on-boarding and remote device management are key requirements. 
In order to carry out supervisory control of the entire system, the \gls{cpps} shall incorporate humans with their cognitive strengths together with machines and their procedural strength \cite{romero_OperatorHumanCyberPhysical_2016}. 
Beyond that, it is necessary to understand the configuration of the system components in order to be able to hierarchically manage them along their lifecycle. 
Exposing management interfaces is another requirement enabling the interaction of industrial facilities and 5G networks.
In \gls{cpps}, the large quantity of data available 
shall be used to let automation components adopt themselves based on the specific context, \ie as location-based service. 
%

\textit{Req-D) Extensibility:} 
A flexible production requires a flexible, extensible and future-proof network technology. 
When services are incrementally added on the factory floor, easy and fast extension of network functions, processing resources and functionalities is required. 
Here, different hierarchies of network, compute and storage resources with interoperable stacks prevent congestion and component redundancy increases reliability. 
The dynamic environments of a flexible production necessitate efficient and flexible ways of smart and autonomous data routing, as well as automatic (edge cloud) application relocation. 
Related to the latter, the system shall provide a lifecycle management of applications and functions to guarantee \gls{qos} for different edge services sharing the same physical resources.

\textit{Req-E) Cyber Security:} 
End-to-end confidentiality and integrity protection mechanisms are key for plant managers, as production and process information belong to their major assets. 
Likewise important is the prevention of fake gNB man-in-the-middle attacks and of mechanisms to purposefully disable certain services, \eg firewalls. 
In addition, strong mechanisms for authentication and authorization of network entities, end devices and their rights have to be provided, which is also associated with a definite identification of subscribers. 
Finally, mechanisms for surveillance and monitoring of sensors, creation of log files for transparent and accountable interaction and audit trail records, \ie chronological documentation of activities that could have affected a particular operation or event during production, are necessary. 

\textit{Req-F) Traffic Separation and Isolation:} 
Manufacturing IT networks are usually segmented for performance and security reasons, in particular according to IEC 62443. 
This becomes an even more important factor in light of \gls{urllc} services and converged networks and necessitates building isolated sub-networks or slices with physical and logical separation of data storage and processing resources for applications that are not interacting with each other or have no common data base. 
Also, setting up inter-site connections with connectivity along logistics routes with trustworthy environments for traffic management at network borders is an important factor.

\section{Reference Architecture Overview}\label{sec_overview}
\begin{figure}
\centering
\vspace{-3mm}
\includegraphics[width=0.8\columnwidth]{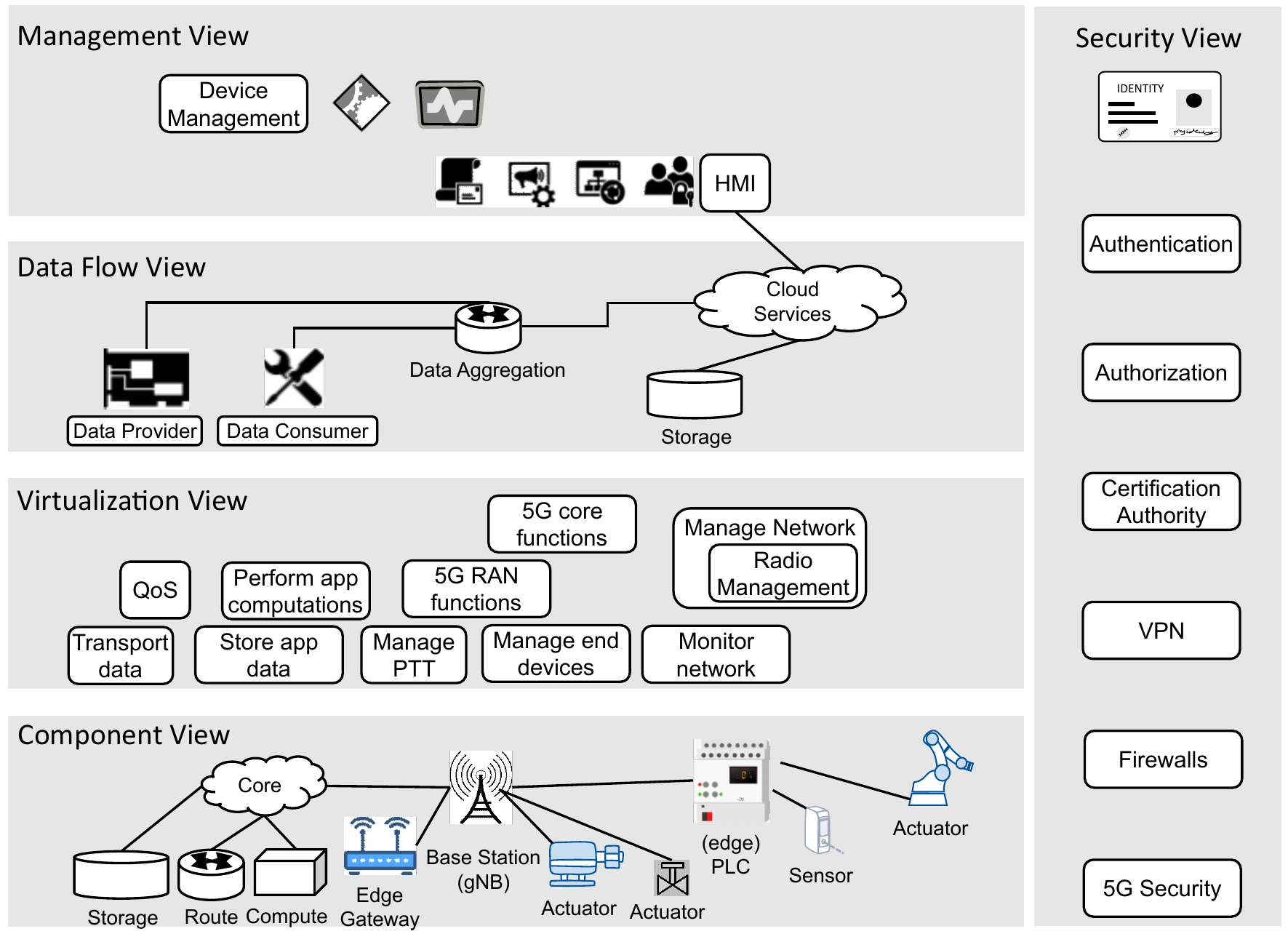}
\caption{Reference Architecture: Overview}
\label{fig_layers}
\end{figure}

\textit{Reference} means that the reference architecture acts as a blueprint for system architectures, which are tailored to actual system requirements. Because plant managers' requirements are that diverse, the reference architecture needs to include a plethora of different views. 
For an overview cf. \cref{fig_layers}, of which present an excerpt of the most important views subsequently. 
The component view represents the physical components with their functionalities, which are abstracted via virtualized components and functions in the virtualization view. 
A third perspective is the data flow view, which describes the routing of generated data from a data provider (\eg sensors) via a data aggregation entity towards a cloud service along the hierarchy levels of the \glsentryshort{rami} model.
Due to space limitations, we omit further details on the data flow view.
The IT security view contains every component and functionality related to the IT security of the architecture and addresses all the other layers while going beyond the application layer. 
Finally, supervisory control and asset management is addressed by the management view.
The presented views of the networking architecture can to a certain extent be projected to the four bottom layers of the \glsentryshort{rami} \cite{adolphs_ReferenceArchitectureModel_2015}, which immediately suggests its practicability. 

\subsection{Component View}\label{sec_component}

\begin{figure}
\centering
\includegraphics[width=0.9\columnwidth]{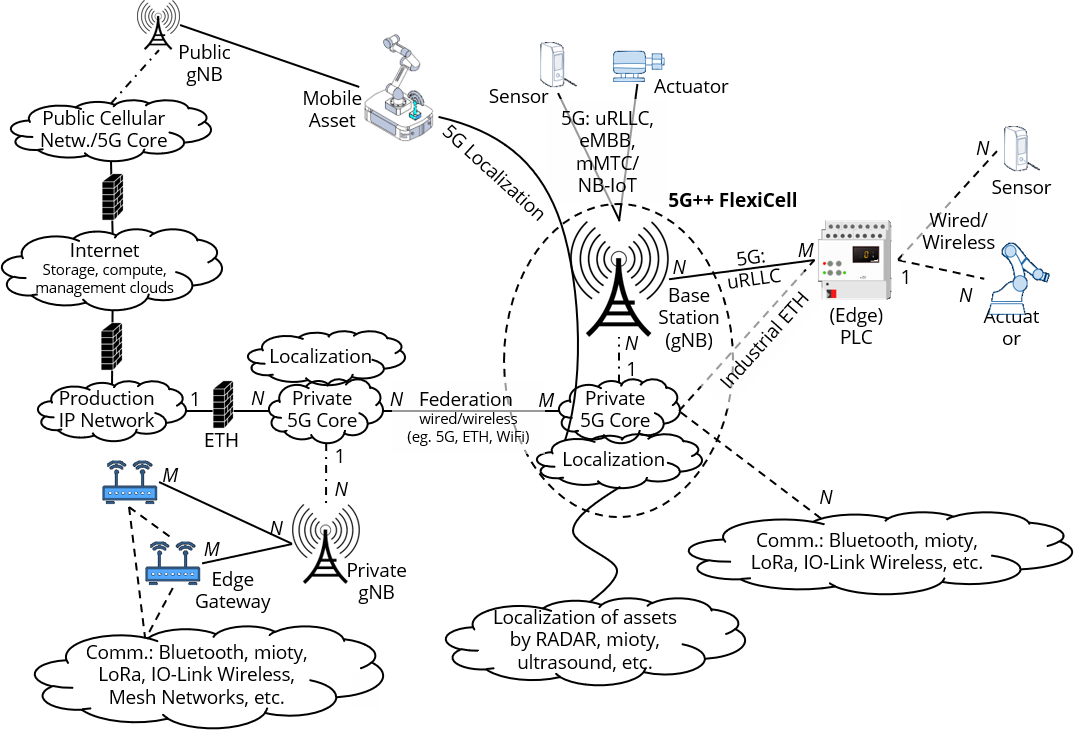}
\caption{Reference Architecture: Component View}
\label{fig_compView}
\end{figure}

The reference architecture includes various components and devices on multiple levels, which have been discussed in earlier contributions and which address many requirements including Req-A, Req-B and Req-F. The interested reader can refer to \cite{ludwig_5G_2018,karrenbauer2019future} for more details.
Of course,  each component can be present in the system multiple times, as the labels connecting them indicate.
Its central builds the 5G++ FlexiCell (\cf \cref{fig_compView}), which is a 5G base stations connected to its local 5G core, such that the FlexiCell can be operated in nomad/island mode, supporting all flavors of 4G/5G (solid lines).
For each production station, where 5G is currently required, a FlexiCell can be set up for local-only coverage. 
The 5G core via \gls{sdn} integrates multiple heterogeneous (dashed lines) wired communication technologies like Industrial Ethernet (\eg Profinet or time-sensitive networking, TSN) classic Ethernet (ETH), field bus networks as well as wireless ones like WiFi, Bluetooth, mioty, LoRa and mesh networks.
Data transmission can be optimized in an \gls{e2e} fashion across different technologies.
Sensors, Actors, (edge) \glspl{plc} can be connected vie 4G/5G or other technologies.
5G edge gateways offer another type of data path, particularly suited for event messaging.
Furthermore, localization technologies like 5G, mioty (via radio frequency), ultrasound or RADAR are integrated into the FlexiCell.
By fusion of different localization technologies position resolution can be improved, and the position data is provided through a technology-agnostic interface, which enables proximity detection and location-based services.

The 5G cores of multiple 5G++ FlexiCells can federate with each other through wired or wireless connections, thus enabling flexible radio coverage with seamless handover.
Via Ethernet connections, FlexiCells can connect to the production IP network and further to the Internet and public cellular networks (all guarded by appropriate IT security measures like firewalls).
Devices might (also in parallel) connect to a public 5G network, if their access rule allow this.
A public--private handover (\eg indoor--outdoor) is possible, if supported by both networks.
The FlexiCell provides a radio management interface (not shown) for \gls{e2e} optimization of heterogeneous data links, see \cref{sec_hmi}

\subsection{Virtualization View}\label{sec_virt}
Resonating with the widespread use of \gls{nfv}, cloud-native architectures and \gls{sdn} in 5G networks \cite{ordonez-lucena_network_2017}, we follow the concept of using commercial-off-the-shelf (COTS) compute, storage and networking hardware. 
While benefiting from low-cost hardware, network functions and edge applications can flexibly and efficiently share the (federated, \cf \cref{sec_component}) hardware resources (Req-D), which could be logically and physically distributed across the factory. 
This is realized by a distributed cloud interconnected through different wireless and wired technologies, which enables effective balancing of resource loads across the factory. The virtualization view 
details how the plant-wide management of physical resources is organized and acts as an intermediate view between the components and the data flow views. 
Along the lines of ETSI reference architectures for \gls{nfv} \cite{ETSI_NFV_002} and \gls{mec} \cite{ETSI_MEC_003}, our architecture consists of three entities, which interact with and manage components, virtual functions and edge applications. 
As an important distinction from the general \gls{nfv} view, we consider sensors and actuators deployed across the plant to be virtualized, as well (\eg as digital twins). 
With this, data provided by massive networks of multipurpose wireless sensors can be shared among different industrial applications facilitating an efficient reuse of machines, robots, etc., reducing costs for plant managers.

Due to space limitations, we omit further details on the Data flow view.

\subsection{Security View} \label{sec_secure}
\begin{figure}
\centering
\vspace{-6mm}
\begin{subfigure}{0.54\textwidth}
	\centering
	\centerline{\includegraphics[width=\columnwidth]{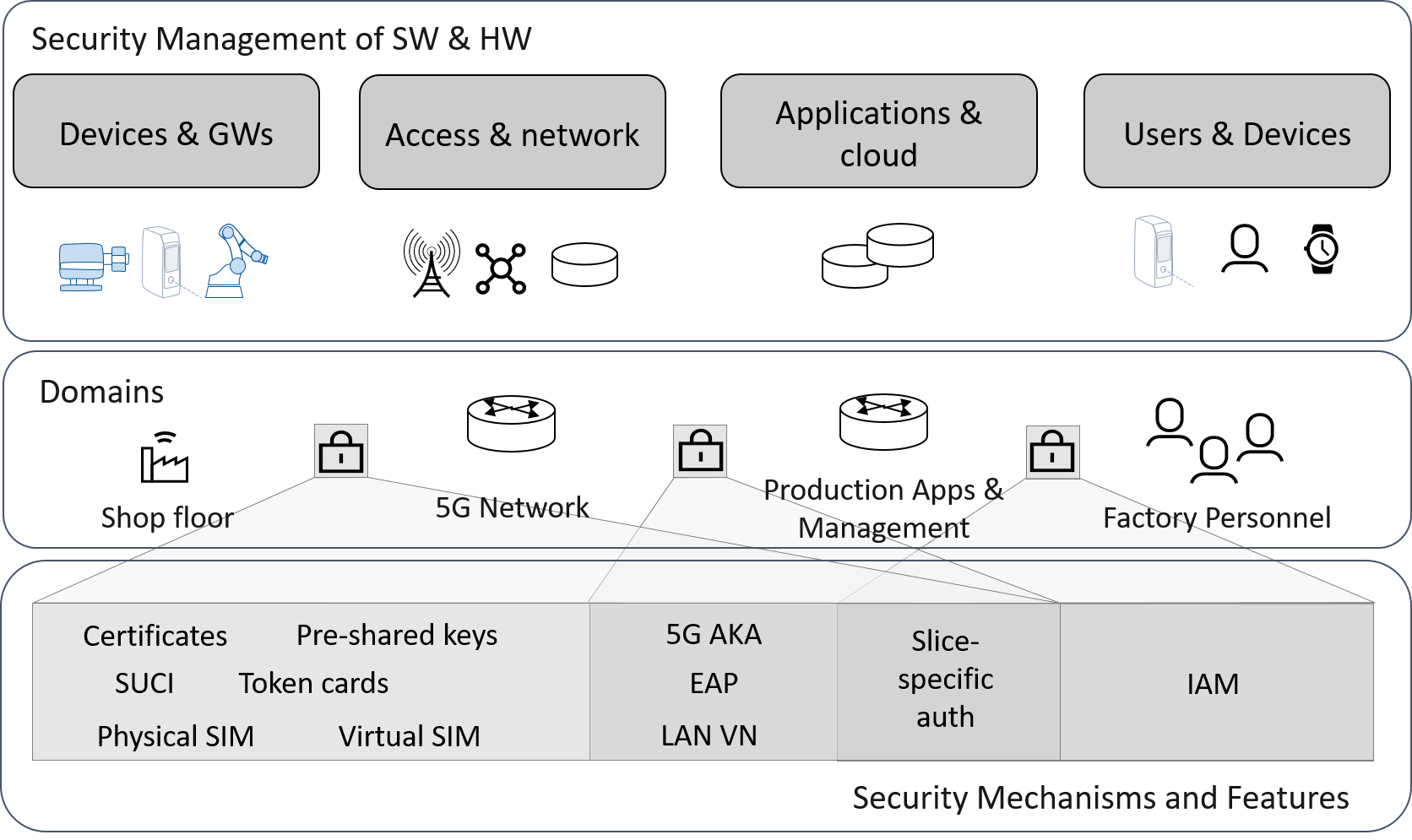}}
	\caption{Security View}
	\label{fig:securityview}
\end{subfigure}
\hfill
\begin{subfigure}{.4\textwidth}
	\centering 
	\includegraphics[width=\textwidth]{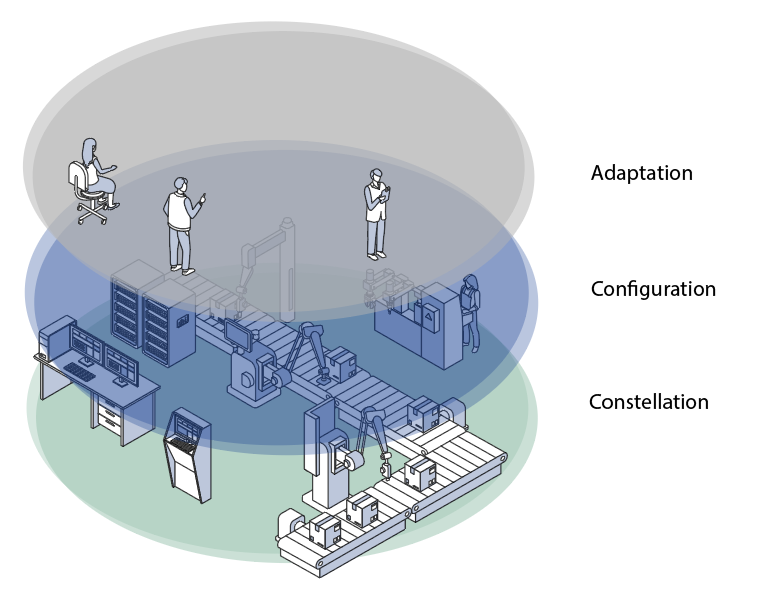}
	\caption{Management View with layers of information}
	\label{fig_MgmtLayers}
\end{subfigure}
\caption{More Reference Architecture Views}\label{fig_ref3}
\end{figure}

The 5G security architecture is designed to provide protection of connected devices  and individuals' privacy (Req-E, Req-F).
Our reference architecture naturally integrates 5G security mechanisms beyond traditional physical SIM cards (\cf \cref{fig:securityview}). 
Physical SIM cards would still continue to be used for mobile broadband subscribers with smartphones, whereas virtual SIM cards are used for \gls{urllc} and \gls{mmtc}. 
Non-SIM-card-based credentials are useful for integration with existing devices (Req-A) and users. 
In the \gls{mno} domain, two alternatives exist, namely the \gls{aka} protocol and the \gls{eap} framework. 
This strong authentication and key agreement procedure also applies for authentication, which the operator delegates to a third party (Req-E). 
By using these mechanisms, components (devices, users and applications) of both worlds can be securely interconnected on different hierarchy levels.

Internal subscriber identification can be done through (e)SIM and digital certificates, and identity and access management systems could be integrated with the 5G system through APIs. 
5G also includes mechanisms to establish trusted network segments with restricted access, among which are LAN virtual networks (VN) and network slicing. 
Especially those address many of the security concerns, including an appropriate coverage of the IEC 62443 and private communication across sites. 
Furthermore, \gls{it}/\gls{ot} integration (management and application) is subject to a study in 3GPP TR 23.745.

\subsection{Management View}\label{sec_hmi}
\begin{figure*}
\centering
\vspace{-3mm}
\includegraphics[width=\columnwidth]{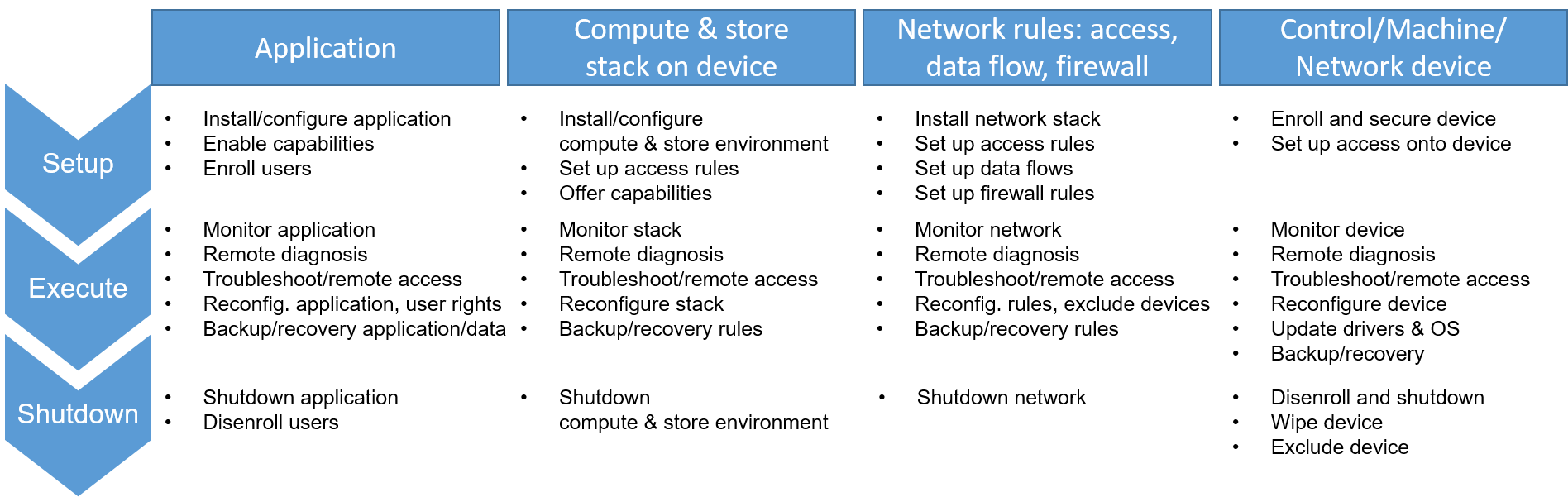}
\vspace{-2mm}
\caption{Reference Architecture: Management View}
\vspace{-4mm}
\label{fig_management}
\end{figure*}

A view on human supervisory and management functions is shown in \cref{fig_management}, which are grouped into device, network function, edge cloud and application functions, on the one hand.
On the other hand, the functions are grouped according to the lifecycle of their (virtual) components (Req-C, Req-D).
For the sake of brevity, we just name some functions and refer to the figure for a longer, albeit non-exhaustive list.
It is clear that not every device will support all functions, but that the management plane is able to cope with the devices' capabilities.
Furthermore, we assume that the management interface supports users in efficiently solving their task with a least amount of knowledge about network technology (Req-A) such that the effort of human interaction is minimized.

The corresponding asset and configuration management is divided into three layers (\cf \cref{fig_MgmtLayers}): 
i) traditional asset management, which keeps the hardware configuration and the location of the devices (\enquote{constellation level});
ii) software configuration of the different assets such that they can be copied to the devices in a container approach. 
For (collaborative) robots, this would be the program code.
iii) we propose the adaptation layer, in which the devices interact with humans.
These layers are considered being orthogonal to those of \gls{it} ecosystems, which apply here too \cite{polenghi_ConceptualModelIT_2020}:
i) operational: governing shop-floor activities, reporting of key performance indicators;
ii) tactical: transformation of long-term objectives to medium/short-term decisions; and 
iii) strategic level: governing an organization's asset portfolio to support capital investment decisions. 
By this, asset and configuration management leverages the ability to generate and manage knowledge \cite{candon_ImplementingIntelligentAsset_2019}, as well as to achieve the balance between performance, costs and risks in a company's business objectives \cite{maletic_LinkAssetRisk_2020}.
Furthermore, it enables self-X features, which keeps complex \gls{cpps} controllable by being self-configurable, self-optimizing, self-healing, self-explanatory and self-protecting \cite{muller2004organic}.

\section{Summary}\label{sec_summary}

Plant managers are still having many concerns and requirements that potentially hinder a widespread roll-out of 5G networks in factories. 
However, many of these requirements can be addressed, when we look at the whole network architecture, instead of considering 5G only as yet another wireless technology. 
In fact, we can see the 5G++ FlexiCell as a kind of glue for the convergence of different already existing communication,  localization technologies. 
We illustrated this in the form of a reference network and localization architecture, which was in parts developed by the project \enquote{5GANG} by the German Federal Ministry of Education and Research, and which is currently built up in the project \enquote{5G-FlexiCell} funded by the Federal Ministry for Economic Affairs and Climate Action. 
Together with company partners, the proposed architecture will be evaluated in three real-life scenarios:
i) a reconfigurable cobot-machine tending in plastic molding; 
ii) a highspeed data sensor on an industrial manipulator in quality control; and 
iii) intra-logistics carried out by a mobile manipulator.

\bibliographystyle{bibtex/splncs_srt} 
\bibliography{5GangArch,reference} 

\end{document}

%% file: acronyms.tex
\newacronym{iot}{IoT}{Internet of Things}
\newacronym{itu}{ITU}{International Telecommunication Union}
\newacronym{iiot}{IIoT}{Industrial Internet of Things}
\newacronym{mc-ss}{MC-SS}{multi-carrier spread spectrum}
\newacronym{mccdma}{MC-CDMA}{multi-carrier code division multiple access}
\newacronym{mmtc}{mMTC}{massive machine type communications}
\newacronym{urllc}{uRLLC}{ultra-reliable and low latency communications}
\newacronym{embb}{eMBB}{enhanced mobile broadband}
\newacronym{sdn}{SDN}{software-defined network}
\newacronym{rru}{RRU}{remote radio unit}
\newacronym{bs}{BS}{base station}
\newacronym{sgw}{S-Gw}{serving gateway}
\newacronym{pgw}{P-Gw}{packet data network gateway}
\newacronym{up}{UP}{user plane}
\newacronym{upf}{UPF}{user plane function}
\newacronym{cp}{CP}{control plane}
\newacronym{ausf}{AUSF}{authentication server function}
\newacronym{udm}{UDM}{user data management}
\newacronym{hss}{HSS}{home subscriber server}
\newacronym{smf}{SMF}{session management function}
\newacronym{mme}{MME}{mobility management entity}
\newacronym{pcf}{PCF}{policy control function}
\newacronym{ap}{AP}{aggregation point}
\newacronym{prp}{PRP}{parallel redundancy protocol}
\newacronym{tcp}{TCP}{transmission control protocol}
\newacronym{api}{API}{application programming interface}
\newacronym{ovs}{OvS}{open vSwitch}
\newacronym{wlan}{WLAN}{wireless local area network}
\newacronym{wan}{WAN}{wide area network}
\newacronym{lpwan}{LPWAN}{low-power wide area network}
\newacronym{ble}{BLE}{Bluetooth low energy}
\newacronym{3gpp}{3GPP}{3rd Generation Partnership Project}
\newacronym{itur}{ITU-R}{International Telecommunication Union Radiocommunication Sector}
\newacronym{qos}{QoS}{quality of service}
\newacronym{nbiot}{NB-IoT}{narrow-band IoT}
\newacronym{ec}{EC}{edge cloud}
\newacronym{nfv}{NFV}{network function virtualization}
\newacronym{iira}{IIRA}{Industrial Internet Reference Architecture}
\newacronym{iic}{IIC}{Industrial Internet Consortium}
\newacronym{rami}{RAMI~4.0}{Reference Architectural Model of Industry 4.0}
\newacronym{mes}{MES}{manufacturing execution system}
\newacronym{erp}{ERP}{enterprise resource planning}
\newacronym{ngmn}{NGMN}{Next Generation Mobile Networks}
\newacronym{agv}{AGV}{automated guided vehicle}
\newacronym{agc}{AGC}{automated guided cart}
\newacronym{slam}{slam}{simultaneous locating and mapping}
\newacronym{mno}{MNO}{mobile network operator}
\newacronym{plc}{PLC}{programmable logic controller}
\newacronym{epc}{EPC}{evolved packet core}
\newacronym{indot}{ind. OT}{industrial operation technology}
\newacronym{ot}{OT}{operation technology}
\newacronym{ict}{ICT}{information and communication technology}
\newacronym{cmtc}{cMTC}{critical Machine Type Communications}
\newacronym{nr}{NR}{New Radio}
\newacronym{das}{DAS}{distributed antenna system}
\newacronym{hmi}{HMI}{human machine interface}
\newacronym{etsi}{ETSI}{European Telecommunications Standards Institute}
\newacronym{mec}{MEC}{multi-access edge computing}
\newacronym{acia}{5G ACIA}{5G Alliance for Connected Industries and Automation}
\newacronym{ue}{UE}{equipment}
\newacronym{hmlv}{HMLV}{high mix, low volume}
\newacronym{sme}{SME}{small and medium-sized enterprise}
\newacronym{5g}{5G}{5th generation of mobile communications}
\newacronym{cpps}{CPPS}{cyber-physical production system}
\newacronym{m2m}{M2M}{machine-to-machine}
\newacronym{roi}{ROI}{return on invest}
\newacronym{it}{IT}{information technology}
\newacronym{e2e}{E2E}{end-to-end}
\newacronym{eap}{EAP}{extensible authentication protocol}
\newacronym{aka}{5G AKA}{5G authentication and key agreement}